\documentclass[12pt,preprint]{aastex}
\usepackage{amsmath,amssymb}
\usepackage{graphicx}
\usepackage{float}

\slugcomment{KSUPT-12/6 \hspace{0.5truecm} December 2012}

\begin{document}
\title{Constraints on dark energy from the Ly${\boldsymbol{\alpha}}$ forest baryon acoustic oscillations measurement of the redshift ${\mathbf{2.3}}$ Hubble parameter}
\author{Omer Farooq and Bharat Ratra}

\affil{Department of Physics, Kansas State University, 
                 116 Cardwell Hall, Manhattan, KS 66506, USA} 
\email{omer@phys.ksu.edu, ratra@phys.ksu.edu}

\begin{abstract}

We use the \cite{Busca12} measurement of the Hubble parameter 
at redshift $z = 2.3$ in conjunction with 21 lower $z$ measurements, 
from \cite{Simon2005}, \cite{gaztanaga09}, \cite{Stern2010}, and 
\cite{moresco12}, to place constraints on model parameters of 
constant and time-evolving dark energy cosmological models. The 
inclusion of the new \cite{Busca12} measurement results in $H(z)$ 
constraints significantly more restrictive than those derived by 
\cite{Farooq2012}. These $H(z)$ constraints are now more restrictive 
than those that follow from current Type Ia supernova (SNIa) apparent 
magnitude measurements \citep{suzuki12}. The $H(z)$ constraints by
themselves require an accelerating cosmological expansion at about 
2 $\sigma$ confidence level, depending on cosmological model and 
Hubble constant prior used in the analysis. A joint analysis of 
$H(z)$, baryon acoustic oscillation peak length scale, and SNIa data 
favors a spatially-flat cosmological model currently dominated by a 
time-independent cosmological constant but does not exclude 
slowly-evolving dark energy density. 

\end{abstract}

\maketitle
\section{Introduction}

Observations indicate the cosmological expansion is accelerating now
and that the Universe is spatially flat, provided the dark energy 
density responsible for the acceleration is close to or time independent.
For reviews of dark energy see \cite{Jimenez2011}, \cite{Li2011a}, 
\cite{Bolotin2011}, \cite{Wang2012}, and references therein.\footnote{
Instead of dark energy that dominates the current cosmological 
energy budget, a less likely possibility is that these observations 
are an indication that general relativity needs to be modified on 
cosmological length scales. See \cite{Bolotin2011}, \cite{Capozziello2011}, 
\cite{Starkman2011}, and references therein, for reviews of 
modified gravity. Here we assume that general relativity 
is an adequate description of cosmological gravitation.}

In the ``standard" spatially-flat $\Lambda$CDM cosmological model
\citep{peebles84} dark energy --- Einstein's cosmological constant $\Lambda$
--- contributes a little more than $70\%$ of the current energy budget.
Non-relativistic cold dark matter (CDM) is the next largest contributer 
(at a little more than $20\%$), followed by non-relativistic baryonic matter 
(around $5\%$). See \cite{Ratra08} and references therein for reviews of 
the standard model. It has been known for a while that the 
standard $\Lambda$CDM model is reasonably compatible with most observational
constraints \citep[see, e.g.,][]{jassal10, wilson06, Davis2007,
allen08}.\footnote{
Note, however, that there are tentative observational indications that 
the ``standard'' CDM structure formation model, which
is assumed in the $\Lambda$CDM model, might need modification 
\citep[e.g.,][]{Peebles&Ratra2003, Perivolaropoulos2010}.} 
In the $\Lambda$CDM model the dark energy density is constant in time and does 
not vary in space.

It is well known that the standard $\Lambda$CDM model has some puzzling 
features that are easier to accept in a model in which the dark energy 
density is a slowly 
decreasing function of time \citep[][]{Peebles&Ratra1988, Ratra&Peebles1988}. 
For recent discussions of time-varying dark energy models see
\cite{Sheykhi2012}, \cite{Brax2012}, \cite{Hollenstein2012}, 
\cite{Cai2012}, \cite{PoncedeLeon2012}, \cite{Costa2012}, \cite{Gu2012},
\cite{Basilakos2012}, \cite{Xu2012}, and references 
therein. In this paper we study two dark energy models (with
dark energy being either a cosmological constant or a slowly-evolving 
scalar field $\phi$) as well as a dark energy parameterization.

In the $\Lambda$CDM model, time-independent dark energy density (the 
cosmological constant $\Lambda$) is modeled as a spatially homogeneous 
fluid with equation of state $p_{\rm \Lambda} = -\rho_{\rm \Lambda}$
that relates the fluid pressure and energy density. The XCDM 
parameterization has often been used to describe slowly-decreasing dark 
energy density. In this case dark energy is modeled as a spatially 
homogeneous $X$-fluid with equation of state $p_{\rm X}=w_{\rm X}
\rho_{\rm X}$. Here $p_{\rm X}$ and $\rho_{\rm X}$ are the pressure 
and energy density of the $X$-fluid and the equation of state parameter 
$w_{\rm X}<-1/3$ is independent of time. When $w_{\rm X}=-1$ the XCDM 
parameterization reduces to the complete, consistent $\Lambda$CDM model. 
For all other values of $w_{\rm X}<-1/3$ the XCDM parameterization 
is incomplete as it does not describe spatial inhomogeneities 
\citep[see, e.g.][]{ratra91, podariu2000}. For computational simplicity, 
in the XCDM case we consider only a spatially-flat cosmological model.  

The $\phi$CDM model is the simplest, complete and consistent model of
slowly-decreasing dark energy density \citep{Peebles&Ratra1988, 
Ratra&Peebles1988}. Here dark energy is modeled as the gradually decreasing 
(in $\phi$ and time) potential energy density $V(\phi)$ of the scalar field. 
In this paper we assume an inverse power-law potential energy 
density $V(\phi) \propto \phi^{-\alpha}$, where $\alpha$ is a nonnegative 
constant \citep{Peebles&Ratra1988}. When $\alpha = 0$ the $\phi$CDM model 
reduces to the corresponding $\Lambda$CDM model. For computational 
simplicity, we assume a spatially-flat cosmology for $\phi$CDM. 

Cosmological observations that provide the strongest evidence for 
dark energy are: SNIa apparent magnitude versus redshift data 
\citep[][and references therein]{suzuki12, Li2011b, 
Astier2012, Ruiz2012}; cosmic microwave background (CMB) anisotropy 
measurements \citep[e.g.,][]{Podariu2001b, Komatsu2011} combined with low 
estimates of the cosmological mass density 
\citep[][and references therein]{chen03b}, provided 
the dark energy density is close to or time independent; and, baryon acoustic 
oscillation (BAO) peak length scale data 
\citep[e.g.,][]{beutler2011, blake11, Mehta2012, Anderson2012}.
Current error bars associated with these three types of 
data are still too large to allow for a significant observational 
discrimination between the $\Lambda$CDM model and the two simple 
time-varying dark energy models discussed above. Additional data 
are needed for this task, as well as to provide a cross
check on the above results.

Other data that have been used for this purpose include
lookback time as a function of redshift \citep[][and references 
therein]{Samushiaetal2010, dantas11, Tonoiu2011, Thakur2012},
gamma-ray burst luminosity distance as a function of redshift 
\citep[e.g.,][]{Samushia&Ratra2010, Wang2011, Busti2012, Poitras2012},
strong gravitational lensing \citep[][and references therein]{chae04, 
lee07, Zhang2010, Cao2012}, HII starburst galaxy 
apparent magnitude as a function of redshift \citep[e.g.,][]{plionisetal10, 
plionisetal11, Mania2012}, angular size as a function of redshift 
\citep[][and references therein]{Chen2012, Lima2012, 
Jackson2012}, and galaxy cluster properties \citep[e.g.,][]{campanelli11,  
Devi2011, DeBoni2012, Gonzalez2012}. The constraints from these data
are, at present, significantly weaker than those from SNIa, BAO, and CMB
anisotropy measurements, but it is anticipated that future data of
these kinds will provide significant constraints.\footnote{
In addition to soon to be available CMB anisotropy data from Planck, 
future space-based SNIa, BAO-like, and galaxy cluster measurements 
\citep[e.g.,][]{podariu01a, Samushia2011, Sartoris2012, Basse2012, 
Majerotto2012, Pavlov2012} should soon provide interesting constraints 
on cosmological parameters.}

Two other current data sets provide interesting constraints on cosmological
parameters, somewhat comparable to those from SNIa, BAO, and CMB anisotropy
data. These are galaxy cluster gas mass fraction as a function of redshift 
measurements \citep[e.g.,][]{allen08, Samushia&Ratra2008, tong11, 
Lu2011, Solano2012} and measurements of the Hubble parameter
as a function of redshift \citep[e.g.,][]{Jimenezetal2003, Samushia&Ratra2006,
samushia07, Sen&Scherrer2008, Chen2011b, Kumar2012, WangZhang2012, Duan2011,
Aviles2012, Seikel2012}. Interestingly, most measurements now provide 
largely compatible constraints on cosmological parameters that are consistent 
with a currently accelerating cosmological expansion. This provides
confidence that the broad outlines of a standard cosmological
model are now in place. 

In this paper we use the 21 $H(z)$ measurements of \cite{Simon2005}, 
\cite{gaztanaga09}, \cite{Stern2010}, and \cite{moresco12}
[listed in Table 1 of \cite{Farooq2012}], in conjunction with 
the $H(z=2.3)$ measurement of \cite{Busca12}, determined from
BAO in the Ly$\alpha$ forest \citep[in combination with WMAP CMB
anisotropy data,][]{Komatsu2011}, to constrain the $\Lambda$CDM and  
$\phi$CDM models and the XCDM parametrization. The inclusion of the 
new \cite{Busca12} measurement results in tighter 
constraints than those derived by \cite{Farooq2012} from the 
21 $H(z)$ measurements alone. The new $H(z)$ constraints derived 
here are more restrictive than those derived from the
recent SNIa data compilation of \cite{suzuki12}, which more carefully
accounts for the systematic errors in SNIa measurements.\footnote{ 
The study of $H(z)$ data is a much less-developed field than that of 
SNIa data, so it is not impossible that future $H(z)$ error bars might 
be larger than what we have used in our analysis here.} 
In addition to deriving 
$H(z)$-data only constraints, we also use these $H(z)$ measurements in 
combination with recent SNIa and BAO data to jointly constrain 
cosmological parameters. Adding the $H(z)$
data tightens the constraints, quite significantly in some parts
of parameter space. More precisely, the $H(z)$ measurements more
significantly tighten constraints on the nonrelativistic matter
density parameter than on the parameter that more closely controls 
the time evolution of the dark energy density

Our paper is organized as follows. In Sec.\ {\ref{equations}} we
present the basic equations of the three dark energy models we
consider. Constraints from the data are derived in Sec.\
{\ref{HzData}}. We conclude in Sec.\ {\ref{summary}}.

\label{intro}

\section{Dark energy models}
\label{equations}

In this section we list relevant characteristics of the two models 
($\Lambda$CDM and $\phi$CDM) and the one parametrization (XCDM)
we use in our analyses of the data.

In the $\Lambda$CDM model with spatial curvature the Hubble parameter
evolves as
\begin{equation}
\label{eq:LCDMFM}
H(z; H_0, \textbf{p}) = H_0 \left[\Omega_{m0} (1+z)^3 + \Omega_{\Lambda}
                          + (1-\Omega_{m0}-\Omega_\Lambda) (1+z)^2 \right]^{1/2} ,
\end{equation}
where $H_0$ is the current value of Hubble parameter
(the Hubble constant), the current value of the spatial curvature density 
parameter is $\Omega_{K0} = 1-\Omega_{m0}-\Omega_{\Lambda}$,
and the model parameter set we want to constrain is 
$\textbf{p}=(\Omega_{m0},\Omega_\Lambda)$. Here $\Omega_{m0}$
is the nonrelativistic (baryonic and cold dark) matter density parameter 
and $\Omega_{\Lambda}$ is the time-independent cosmological constant 
density parameter. Below we shall have need for the dimensionless 
Hubble parameter $E(z) = H(z)/H_0$. 

The XCDM parameterization Friedmann equation is
\begin{equation}
   H(z; H_0, \textbf{p}) = H_0 [\Omega_{m0}(1+z)^3 + 
      (1 - \Omega_{m0}) (1+z)^{3(1+\omega_{\rm X})}]^{1/2}, 
\end{equation}
where for computational simplicity we consider only flat spatial 
hypersurfaces and the model parameters $\textbf{p}= (\Omega_{m0} ,
\omega_{\rm X})$. The XCDM parametrization is incomplete, as it
cannot describe the evolution of energy density inhomogeneities.

The $\phi$CDM model \citep{Peebles&Ratra1988} is the 
simplest, complete and consistent dynamical dark energy model. 
In this model dark energy is a slowly-rolling scalar field 
$\phi$ with an, e.g., inverse-power-law potential energy density 
$V(\phi)=\kappa m_p^2 \phi^{-\alpha} $ where $m_p=1/\sqrt{G}$ is 
the Planck mass, $G$ is the Newtonian gravitational constant, and 
$\alpha$ is a non-negative free parameter that determines $\kappa$. 
The scalar field part of the $\phi$CDM model action is
\begin{equation}
   S=\frac{m_p^2}{16\pi}\int{\sqrt{-g}\left( \frac{1}{2} ~g^{\mu \nu}
   \partial_\mu \phi \partial_\nu \phi - \kappa m_p^2 \phi^{-\alpha} 
   \right) d^4x},
\end{equation}
where $g^{\mu\nu}$ is the metric tensor and $\alpha$ and $\kappa$ are related as
\begin{equation}
\kappa = \frac{8}{3}\left(\frac{\alpha+4}{\alpha+2}\right)\left(\frac{2 \alpha (\alpha+2)}{3}\right)^{\alpha /2}, 
\end{equation}
with corresponding scalar field equation of motion 
\begin{equation}
\label{eq:dotphi} 
    \ddot{\phi} + 3 \frac{\dot{a}}{a}\dot{\phi} -
    \kappa \alpha m_p^2 \phi^{-(\alpha+1)} = 0,
\end{equation}
where the overdot denotes a derivative with respect to time. 
In the spatially-flat case the Friedmann equation for the $\phi$CDM model is
\begin{equation}
\label{eq:phicdmfriedman}
   H(z; H_0, \textbf{p}) = H_0[\Omega_{m0}(1+z)^3+\Omega_\phi 
   (z,\alpha)]^{1/2},
\end{equation}
where $\Omega_\phi(z, \alpha)$ is determined by the $\phi$ field energy density
\begin{eqnarray}
\label{eq:rhom}
   \rho_\phi = \frac{m_p^2}{16\pi} \left({\frac{1}{2}}\dot{\phi}^2 
   + \kappa m_p^2 \phi^{-\alpha}\right).
\end{eqnarray}
Equations ({\ref{eq:dotphi}})---({\ref{eq:rhom}}) constitute a system of
differential equations which can be solved numerically for the $\phi$CDM 
model Hubble parameter $H(z)$, using the initial conditions described in 
\cite{Peebles&Ratra1988}. In this case the model parameter 
set is $\textbf{p}=(\Omega_{m0},\alpha)$.

\section{Constraints from the data}
\label{HzData}

\begin{figure}[t]
\centering
  \includegraphics[width=80mm]{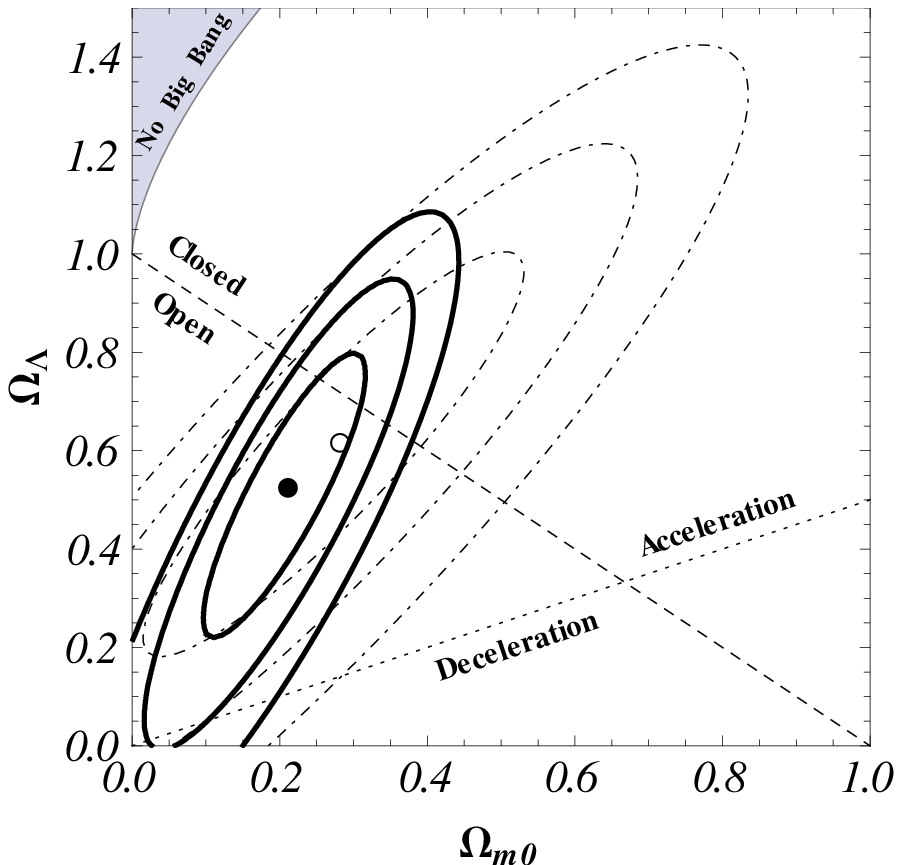}
  \includegraphics[width=80mm]{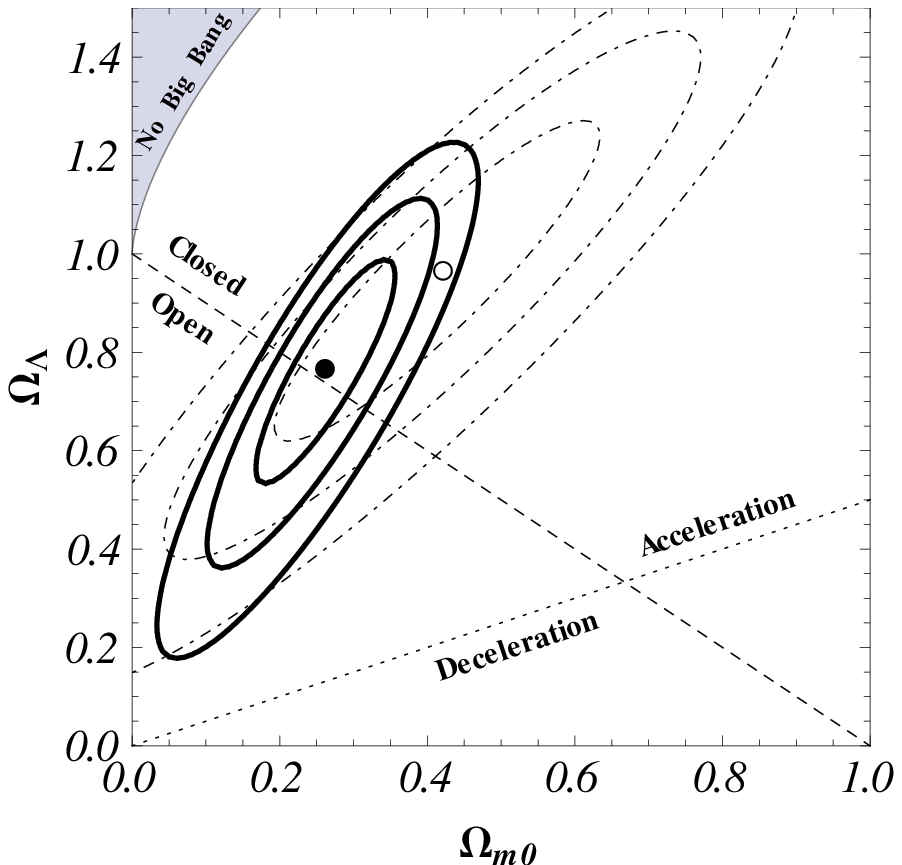}

\caption{
Thick solid (thin dot-dashed) lines correspond to 1, 2, and 3 $\sigma$ 
constraint contours from the new \citep[old,][]{Farooq2012} $H(z)$ data 
for the $\Lambda$CDM model. The filled (empty) circle is the best fit 
point from the new (old) $H(z)$ data. The left panel is for the 
$H_0 = 68 \pm 2.8$ km s$^{-1}$ Mpc$^{-1}$ prior and the right 
panel is for the $H_0 = 73.8 \pm 2.4$ km s$^{-1}$ Mpc$^{-1}$ one.
The dashed diagonal lines correspond to spatially-flat models, the 
dotted lines demarcate zero-acceleration models, and the shaded area 
in the upper left-hand corners are the region  for which there is no 
big bang. The filled circles correspond to best-fit pair 
$(\Omega_{m0}, \Omega_{\Lambda}) = (0.21, 0.53)$ with $\chi^2_{\rm min}
= 15.1$ (left panel) and best-fit pair $(\Omega_{m0}, \Omega_{\Lambda}) = 
(0.26, 0.77)$ with $\chi^2_{\rm min}=16.1$ (right panel). The empty 
circles correspond to best-fit pair 
$(\Omega_{m0}, \Omega_{\Lambda}) = (0.28, 0.62)$ with $\chi^2_{\rm min}
= 14.6$ (left panel) and best-fit pair $(\Omega_{m0}, \Omega_{\Lambda}) = 
(0.42, 0.97)$ with $\chi^2_{\rm min}=14.6$ (right panel). 
} \label{fig:LCDM_Hz}
\end{figure}


\begin{figure}[t]
\centering
  \includegraphics[width=80mm]{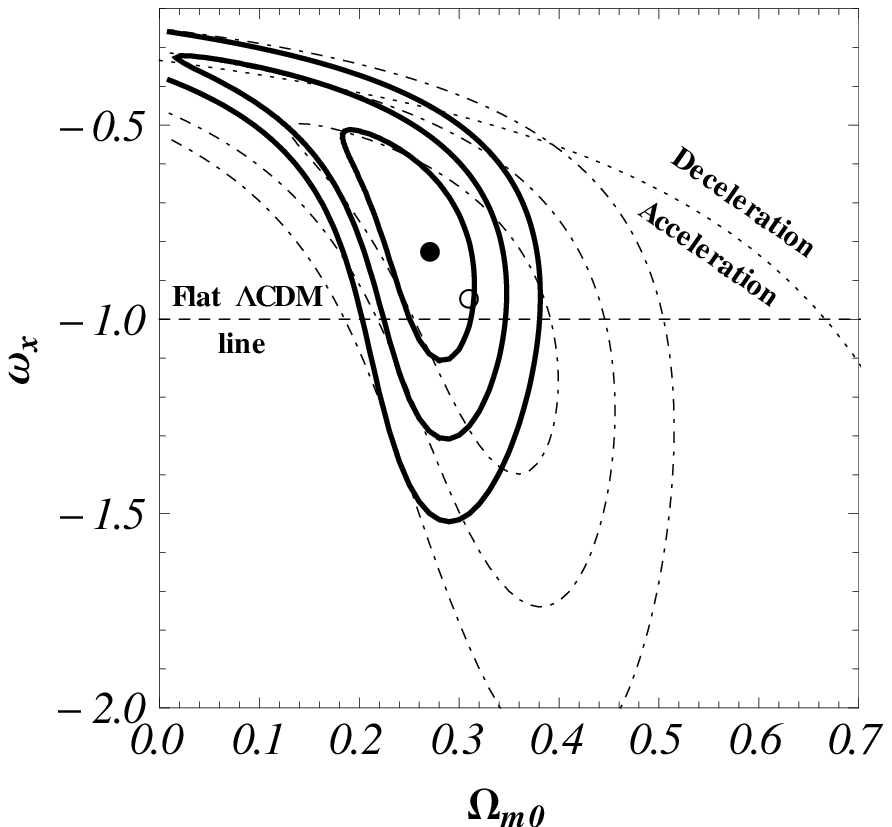}
  \includegraphics[width=80mm]{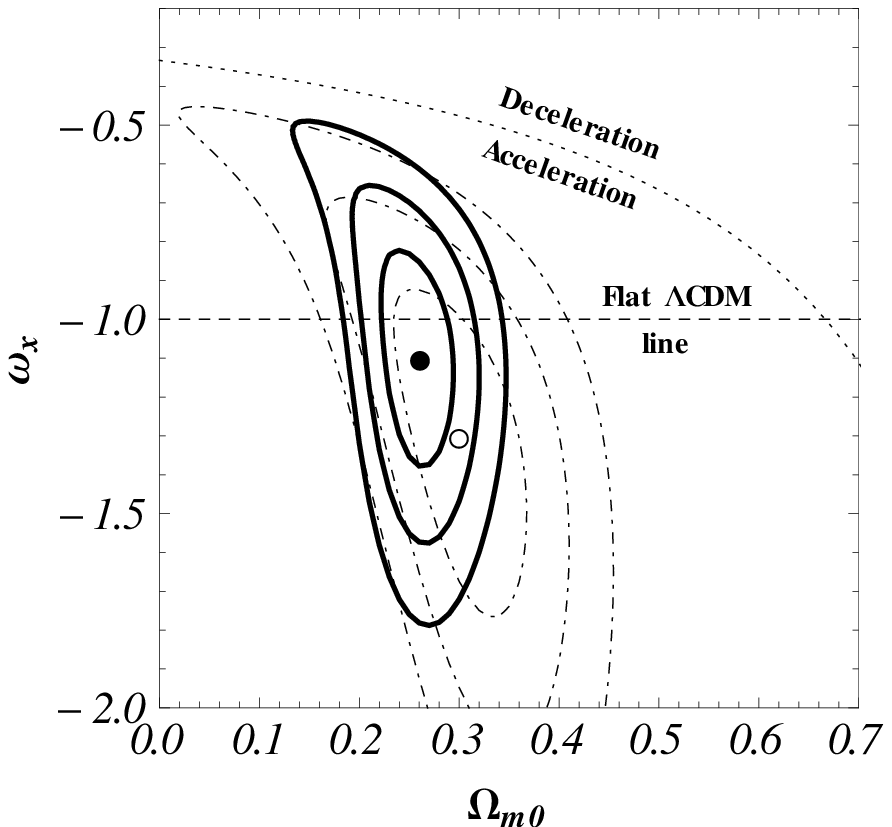}

\caption{
Thick solid (thin dot-dashed) lines correspond to 1, 2, and 3 $\sigma$ 
constraint contours from the new \citep[old,][]{Farooq2012} $H(z)$ data 
for the XCDM model. The filled (empty) circle is the best fit point from 
the new (old) $H(z)$ data. The left panel is for the 
$H_0 = 68 \pm 2.8$ km s$^{-1}$ Mpc$^{-1}$ prior and the right 
panel is for the $H_0 = 73.8 \pm 2.4$ km s$^{-1}$ Mpc$^{-1}$ one.
The dashed horizontal lines at $\omega_{\rm X} = -1$ correspond to 
spatially-flat $\Lambda$CDM models and the curved dotted lines demarcate 
zero-acceleration models. The filled circles correspond to best-fit 
pair $(\Omega_{m0}, \omega_{\rm X}) = (0.27, -0.82)$ with $\chi^2_{\rm min}
= 15.2$ (left panel) and best-fit pair $(\Omega_{m0}, \omega_{\rm X}) 
= (0.36, -1.1)$ with $\chi^2_{\rm min}=15.9$ (right panel).
The empty circles correspond to best-fit 
pair $(\Omega_{m0}, \omega_{\rm X}) = (0.31, -0.94)$ with $\chi^2_{\rm min}
= 14.6$ (left panel) and best-fit pair $(\Omega_{m0}, \omega_{\rm X}) 
= (0.30, -1.30)$ with $\chi^2_{\rm min}=14.6$ (right panel).
} \label{fig:XCDM_Hz}
\end{figure}

\begin{figure}[t]
\centering
  \includegraphics[width=80mm]{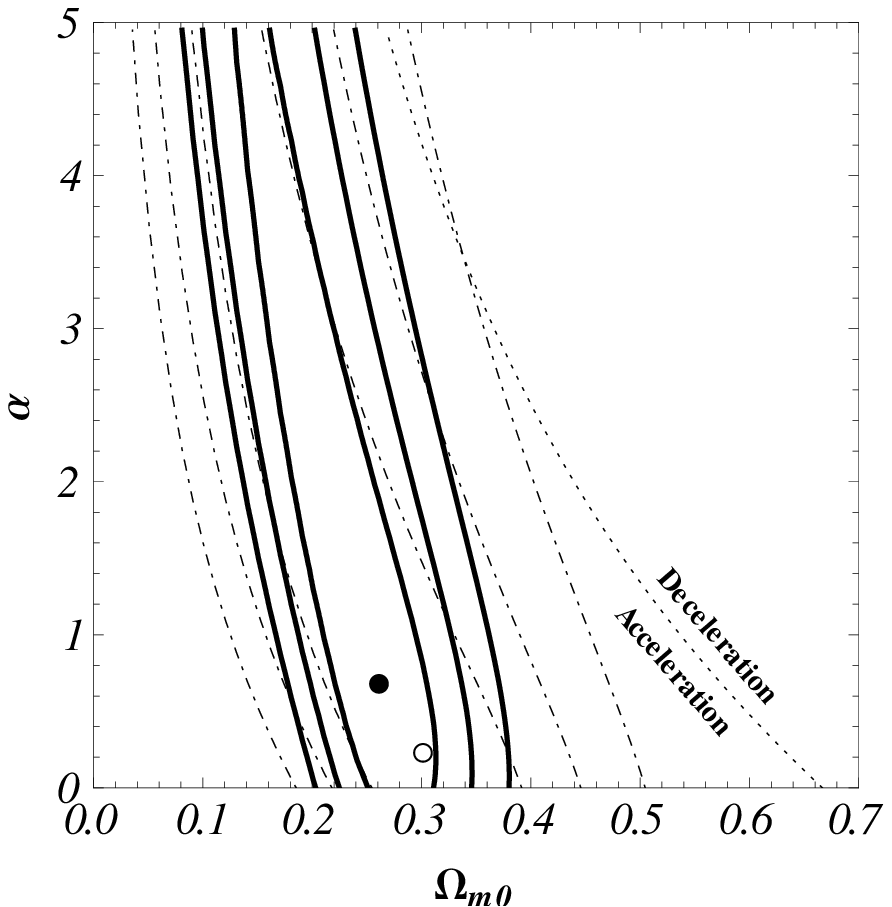}
  \includegraphics[width=80mm]{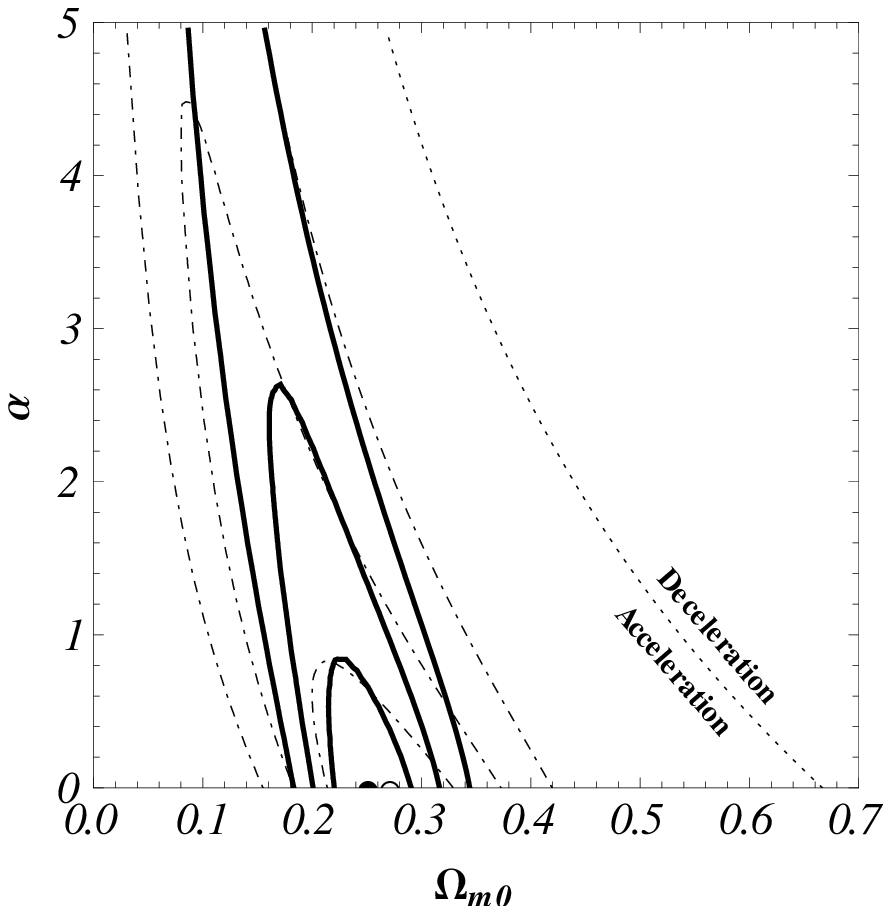}

\caption{
Thick solid (thin dot-dashed) lines correspond to 1, 2, and 3 $\sigma$ 
constraint contours from the new \citep[old,][]{Farooq2012} $H(z)$ data 
for the $\phi$CDM model. The filled (empty) circle is the best fit point 
from the new (old) $H(z)$ data. The left panel is for the 
$H_0 = 68 \pm 2.8$ km s$^{-1}$ Mpc$^{-1}$ prior and the right 
panel is for the $H_0 = 73.8 \pm 2.4$ km s$^{-1}$ Mpc$^{-1}$ one.
The horizontal axes at $\alpha = 0$ correspond to spatially-flat 
$\Lambda$CDM models and the curved dotted lines demarcate 
zero-acceleration models. The filled circles correspond to best-fit pair 
$(\Omega_{m0}, \alpha) = (0.36, 0.70)$ with $\chi^2_{\rm min}=15.2$
(left panel) and best-fit pair $(\Omega_{m0}, \alpha) = (0.25, 0)$ 
with $\chi^2_{\rm min}=16.1$ (right panel). The empty circles correspond to best-fit pair 
$(\Omega_{m0}, \alpha) = (0.30, 0.25)$ with $\chi^2_{\rm min}=14.6$
(left panel) and best-fit pair $(\Omega_{m0}, \alpha) = (0.27, 0)$ 
with $\chi^2_{\rm min}=15.6$ (right panel).
} \label{fig:phiCDM_Hz}
\end{figure}

We first study $H(z)$ data constraints on cosmological parameters.
To the 21 independent $H(z)$ data points listed in Table 1 of 
\cite{Farooq2012} we add the \cite{Busca12} $H(z = 2.3) = 224 \pm 8$
km s$^{-1}$ Mpc$^{-1}$ measurement, determined from BAO in the 
Ly$\alpha$ forest 
\citep[in conjunction with WMAP CMB anisotropy data,][]{Komatsu2011}. 

To constrain cosmological parameters $\textbf{p}$ of the models of 
interest we follow the procedure of \cite{Farooq2012}. We again 
marginalize over the nuisance parameter $H_0$ using two different 
Gaussian priors with $\bar{H_0}\pm\sigma_{H_0}$=
68 $\pm$ 2.8 km s$^{-1}$ Mpc$^{-1}$ and with $\bar{H_0}\pm\sigma_{H_0}$=
73.8 $\pm$ 2.4 km s$^{-1}$ Mpc$^{-1}$. As discussed there, the Hubble 
constant measurement uncertainty can significantly affect 
cosmological parameter estimation 
\citep[for a recent example see, e.g.,][]{calabrese12}. 
The lower of the two values we use is from a median 
statistics analysis \citep{Gott2001} of 553 measurements of $H_0$
\citep{Chen2011a}; this estimate has been stable for over 
a decade now \citep{Gott2001, Chen2003}. The other value is a recent, 
HST based one \citep{Riess2011}. Other recent estimates are 
compatible with at least one of the two values we use 
\citep[see, e.g.,][]{Freedman2012, Sorce2012, Colless2012, Tammann2012}. 

We maximize the likelihood $\mathcal{L}_H(\textbf{p})$ with respect 
to the parameters $\textbf{p}$ to find the best-fit parameter values 
$\mathbf{p_0}$. In the models we consider $\chi_H^2 = -2 {\rm ln}
[\mathcal{L}_H(\textbf{p})] $ depends on 
two parameters. We define 1, 2, and 3 $\sigma$ 
confidence intervals as two-dimensional parameter sets bounded by 
$\chi_H^2(\textbf{p}) = \chi_H^2(\mathbf{p_0})+2.3,~\chi_H^2(\textbf{p}) = 
\chi_H^2(\mathbf{p_0})+6.17$, and $\chi_H^2(\textbf{p}) = 
\chi_H^2(\mathbf{p_0})+11.8$, respectively.

Figures \ref{fig:LCDM_Hz}---\ref{fig:phiCDM_Hz} show the constraints
from the $H(z)$ data derived here, as well as those derived by 
\cite{Farooq2012}, for the three dark energy models, and for the two 
different $H_0$ priors. Clearly, the $H(z = 2.3)$ measurement of 
\cite{Busca12} significantly tightens the constrains. Given that the 
nonrelativistic matter density is larger relative to the dark energy density 
at $z = 2.3$, it is perhaps not unexpected that the \cite{Busca12} 
measurement tightens the constraints on $\Omega_{m0}$ much more 
significantly than it does for the constraints on the other 
parameter which more strongly affects the evolution of the dark energy
density, see Figs.\ \ref{fig:XCDM_Hz} and \ref{fig:phiCDM_Hz}.

\begin{figure}[t]
\centering
  \includegraphics[angle=0,width=80mm]{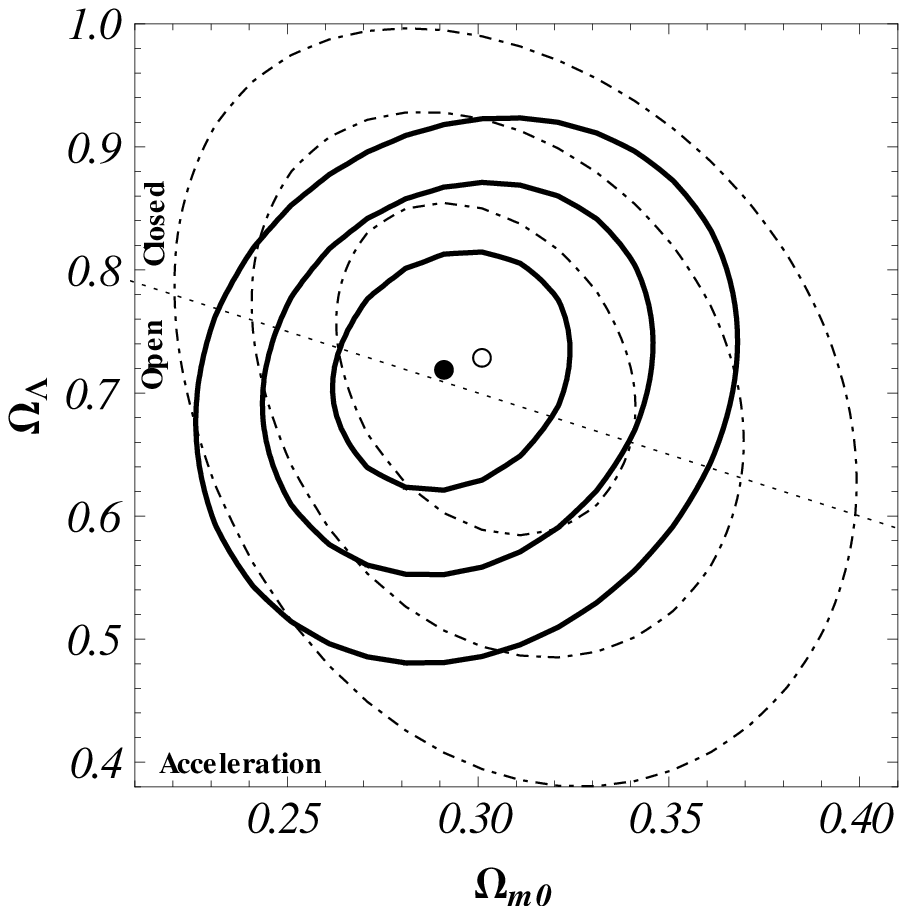}
  \includegraphics[angle=0,width=80mm]{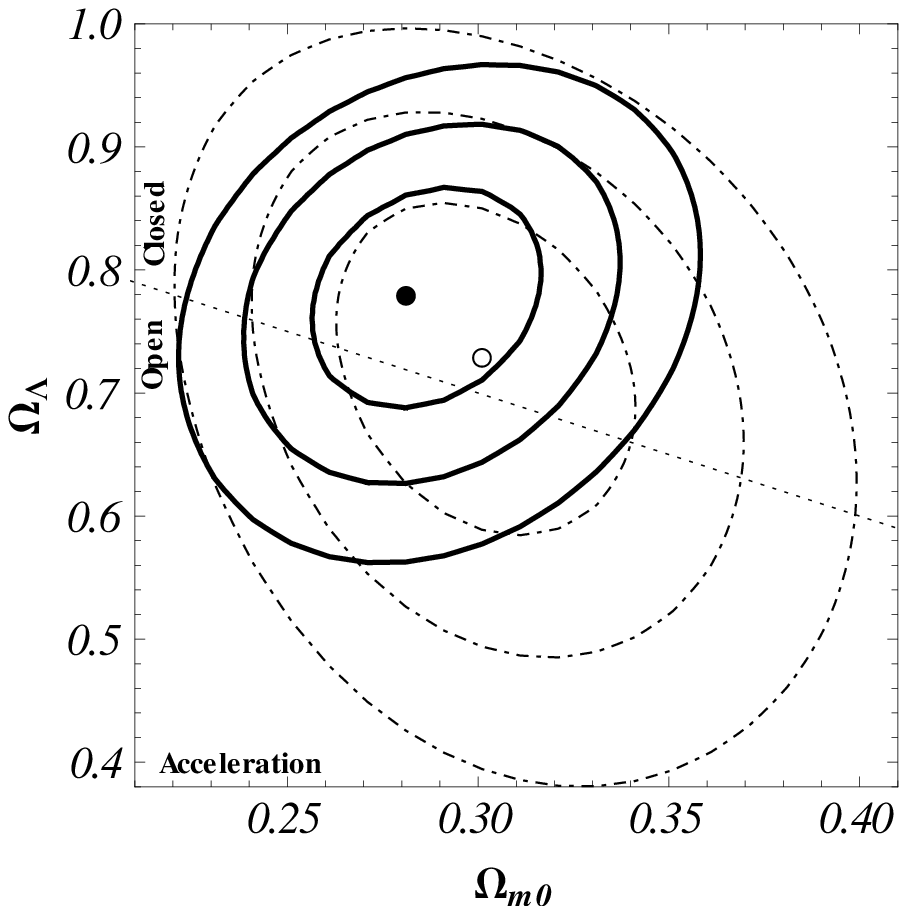}
\caption{
Thick solid (thin dot-dashed) lines are 1, 2, and 3 $\sigma$ constraint
contours for the $\Lambda$CDM model from a joint analysis of the BAO
and SNIa (with systematic errors) data, with (without) the $H(z)$ data. 
The full (empty) circle marks the best-fit point determined from the 
joint analysis with (without) the $H(z)$ data. The dotted sloping line 
corresponds to spatially-flat $\Lambda$CDM models. In the left panel 
we use the $H_0$ = 68 $\pm$ 2.8 km s$^{-1}$ Mpc$^{-1}$ prior. Here the
empty circle [no $H(z)$ data] corresponds to best-fit pair 
$(\Omega_{m0}, \Omega_{\Lambda}) = (0.30,0.73)$ with $\chi^2_{\rm min}=551$
while the full circle [with $H(z)$ data] indicates best-fit pair 
$(\Omega_{m0}, \Omega_{\Lambda}) = (0.29,0.72)$ with $\chi^2_{\rm min}=567$. 
In the right panel we use the $H_0$ = 73.8 $\pm$ 2.4 km s$^{-1}$ Mpc$^{-1}$ 
prior. Here the empty circle [no $H(z)$ data] corresponds to best-fit pair 
$(\Omega_{m0}, \Omega_{\Lambda}) = (0.30,0.73)$ with $\chi^2_{\rm min}=551$
while the full circle [with $H(z)$ data] demarcates best-fit pair 
$(\Omega_{m0}, \Omega_{\Lambda}) = (0.28,0.78)$ with $\chi^2_{\rm min}=568$.
}
\label{fig:LCDM_com}
\end{figure}

\begin{figure}[t]
\centering
  \includegraphics[angle=0,width=80mm]{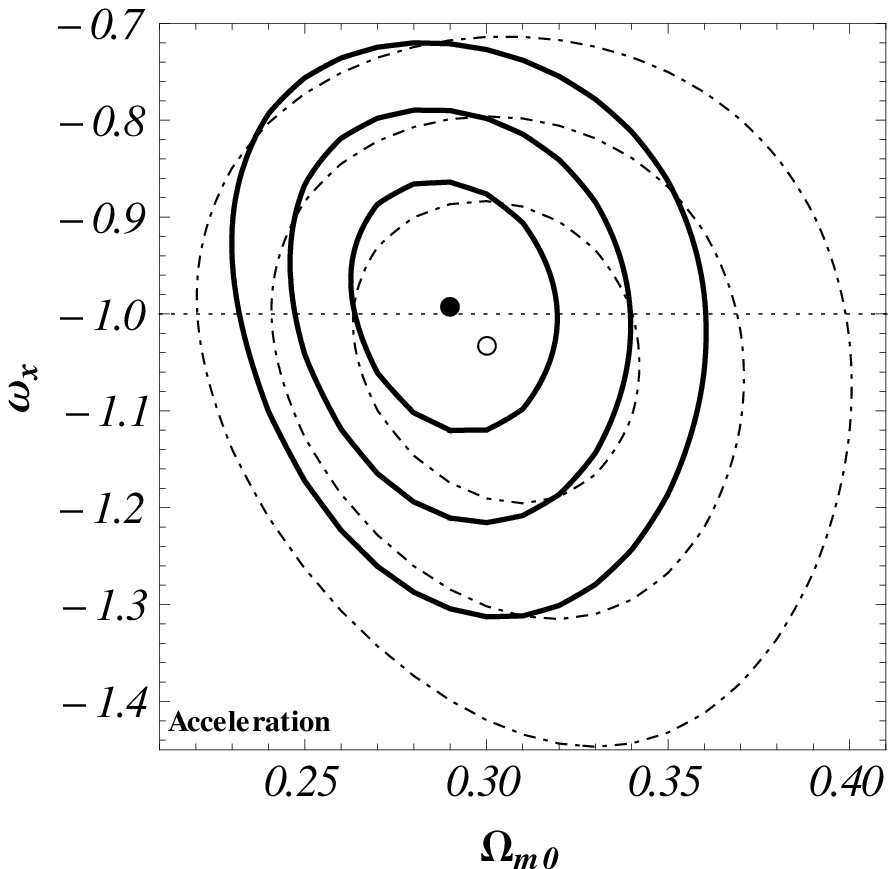}
  \includegraphics[angle=0,width=80mm]{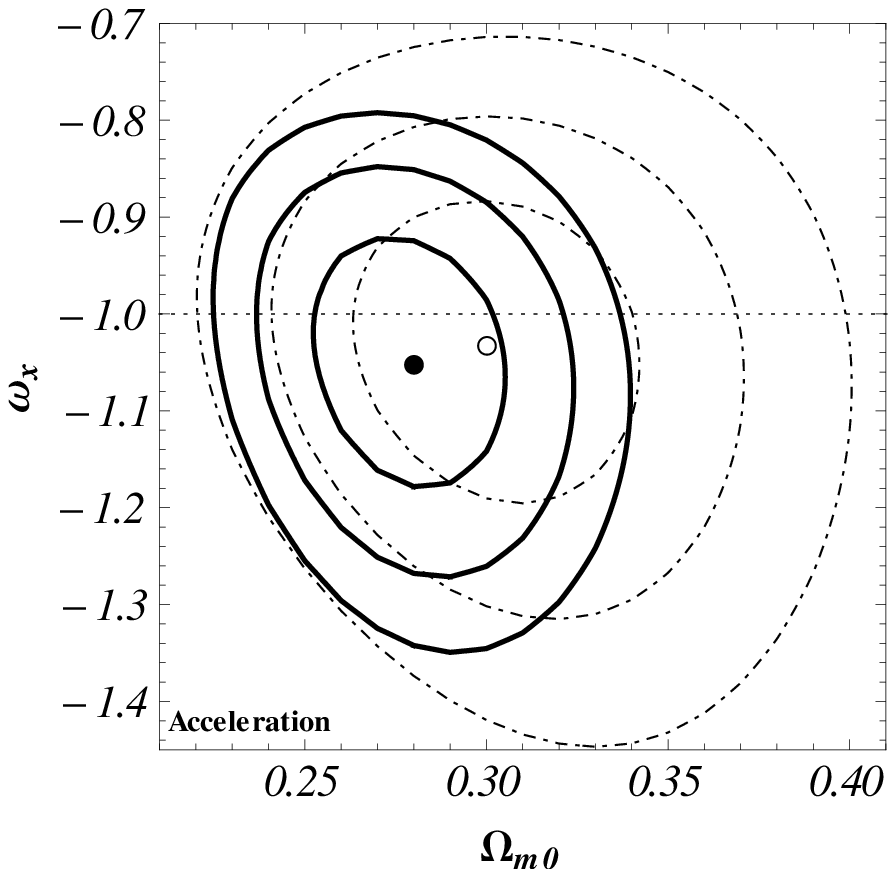}
 \caption{
Thick solid (thin dot-dashed) lines are 1, 2, and 3 $\sigma$ constraint
contours for the XCDM parametrization from a joint analysis of the BAO
and SNIa (with systematic errors) data, with (without) the $H(z)$ data. 
The full (empty) circle marks the best-fit point determined from the 
joint analysis with (without) the $H(z)$ data. The dotted horizontal 
line at $\omega_{\rm X} =-1$  corresponds to spatially-flat $\Lambda$CDM 
models. In the left panel we use the 
$H_0$ = 68 $\pm$ 2.8 km s$^{-1}$ Mpc$^{-1}$ prior. Here the empty
circle [no $H(z)$ data] corresponds to best-fit pair $(\Omega_{m0}, 
\omega_{\rm X}) = (0.30,-1.03)$ with $\chi^2_{\rm min}=551$, while 
the full circle [with $H(z)$ data] demarcates best-fit pair 
$(\Omega_{m0}, \omega_{\rm X}) = (0.29,-0.99)$ with $\chi^2_{\rm min}=568$. 
In the right panel we use the $H_0$ = 73.8 $\pm$ 2.4 km s$^{-1}$ Mpc$^{-1}$ 
prior. Here the empty circle [no $H(z)$ data] corresponds to best-fit pair 
$(\Omega_{m0}, \omega_{\rm X}) = (0.30,-1.03)$ with $\chi^2_{\rm min}=551$
while the full circle [with $H(z)$ data] indicates best-fit pair 
$(\Omega_{m0}, \omega_{\rm X}) = (0.28,-1.05)$ with $\chi^2_{\rm min}=569$. 
} \label{fig:XCDM_com}
\end{figure}

\begin{figure}[t]
\centering
  \includegraphics[angle=0,width=80mm]{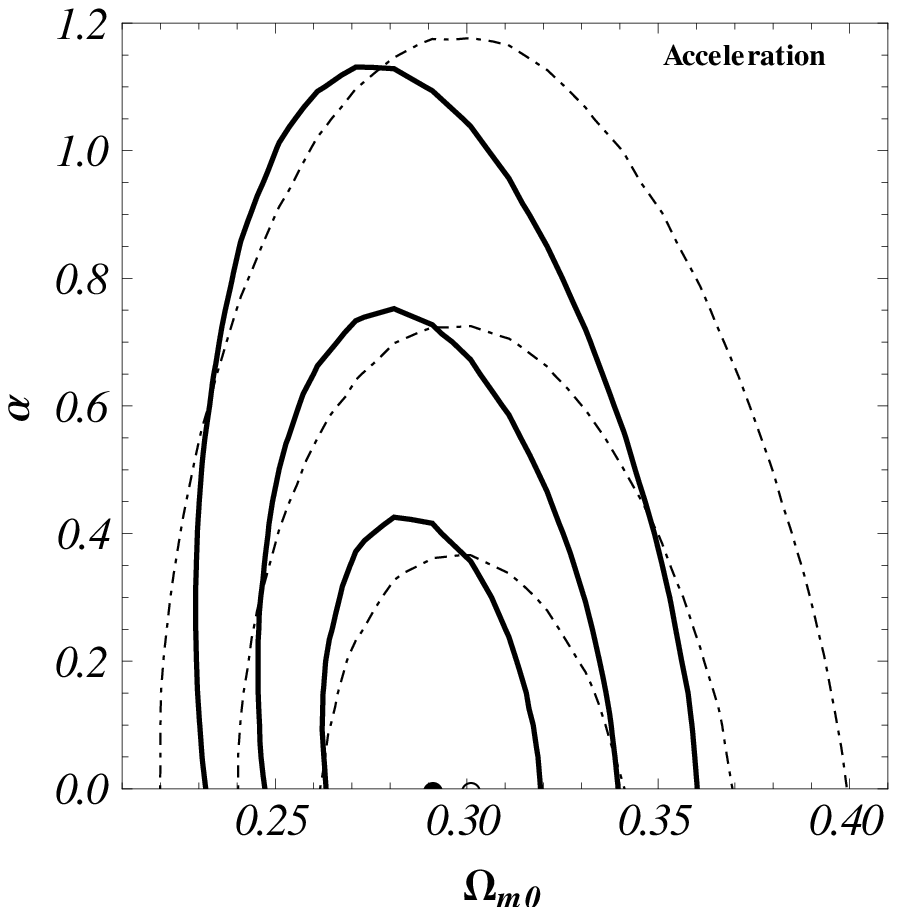}
  \includegraphics[angle=0,width=80mm]{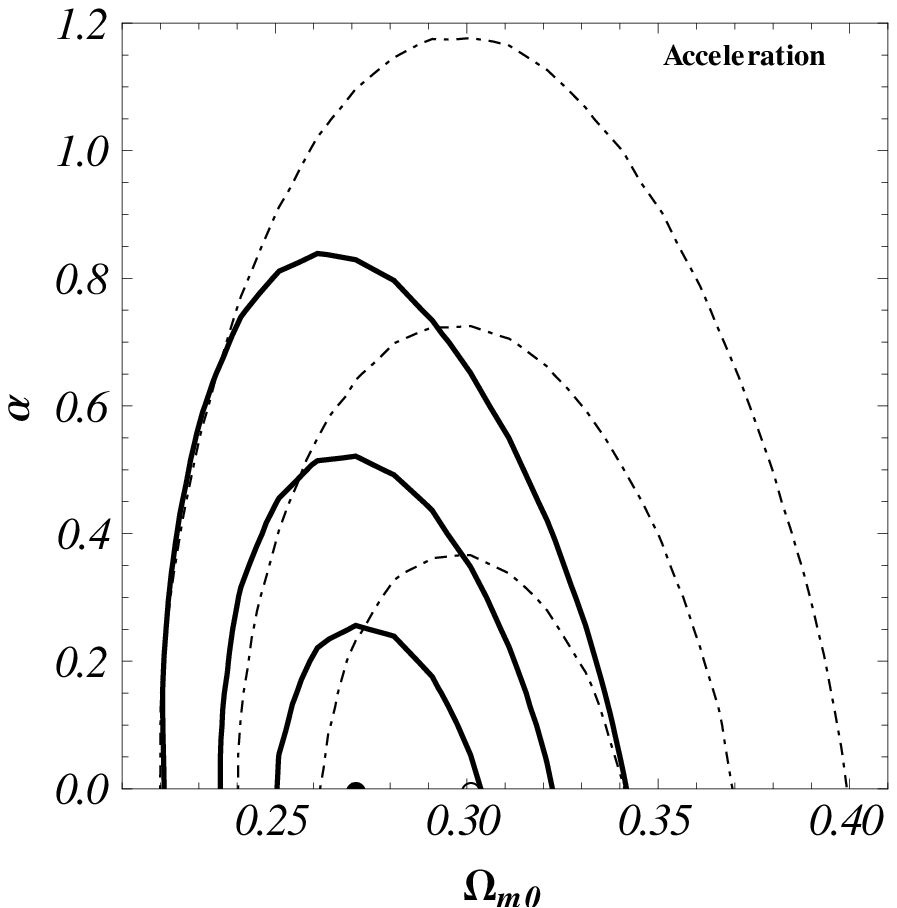}

 \caption{
Thick solid (thin dot-dashed) lines are 1, 2, and 3 $\sigma$ constraint
contours for the $\phi$CDM model from a joint analysis of the BAO
and SNIa (with systematic errors) data, with (without) the $H(z)$ 
data. The full (empty) circle marks the best-fit point determined
from the joint analysis with (without) the $H(z)$ data. The $\alpha = 0$
horizontal axes correspond to spatially-flat $\Lambda$CDM models.
In the left panel we use the $H_0$ = 68 $\pm$ 2.8 km s$^{-1}$ Mpc$^{-1}$ 
prior. Here the empty circle corresponds to best-fit pair 
$(\Omega_{m0}, \alpha) = (0.30, 0)$ with $\chi^2_{\rm min}=551$
while the full circle indicates best-fit pair $(\Omega_{m0}, \alpha) 
= (0.29, 0)$ with $\chi^2_{\rm min}=567$. In the right panel we use 
the $H_0$ = 73.8 $\pm$ 2.4 km s$^{-1}$ Mpc$^{-1}$ prior. Here the empty
circle corresponds to best-fit pair $(\Omega_{m0}, \alpha) = (0.30, 0)$ 
with $\chi^2_{\rm min}=551$ while the full circle demarcates best-fit 
pair $(\Omega_{m0}, \alpha) = (0.27, 0)$ with $\chi^2_{\rm min}=569$.
}
\label{fig:phiCDM_com}
\end{figure}

Comparing the $H(z)$ constraints derived here, and shown in Figs.\ 
\ref{fig:LCDM_Hz}---\ref{fig:phiCDM_Hz} here, to the SNIa constraints
shown in Fig.\ 4 of \cite{Farooq2012}, we see that the new $H(z)$
data constraints are significantly more restrictive than those 
that follow on using the SNIa data. This is a remarkable result.
Qualitatively, because of the dependence on the $H_0$ prior 
and on the model used in the analysis, Figs.\ 
\ref{fig:LCDM_Hz}---\ref{fig:phiCDM_Hz} show that the $H(z)$
data alone require accelerated cosmological expansion at 
approximately the two standard deviation confidence level.

While the $H(z)$ data provide tight constraints on a linear
combination of cosmological parameters, the banana-like constraint
contours of Figs.\ \ref{fig:LCDM_Hz}---\ref{fig:phiCDM_Hz}
imply that these data alone cannot significantly discriminate
between cosmological models. To tighten the constraints we must add other 
data to the mix. Following \cite{Farooq2012}, we derive constraints on 
cosmological parameters of the three models from a joint analysis of the $H(z)$
data with the 6 BAO peak length scale measurements of \cite{Percival2010},
\cite{beutler2011}, and \cite{blake11}, and the Union2.1 compilation of
580 SNIa apparent magnitude measurements (covering a redshift range
$0.015<z <1.4$) from \cite{suzuki12}.

\begin{figure}[t]
\centering
  \includegraphics[angle=0,width=160mm]{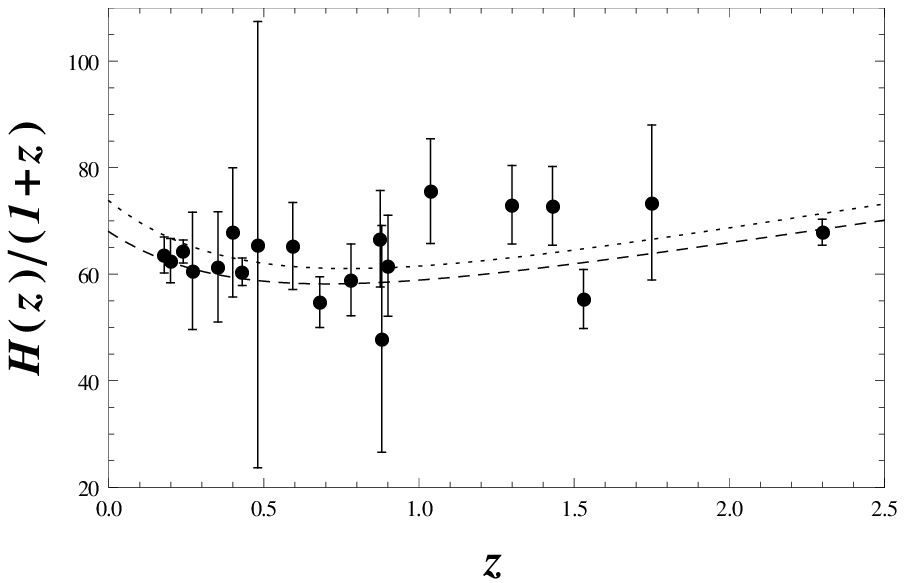}

\caption{
Measurements and predictions for $H(z)/(1+z)$ as a function of $z$.
Dashed (dotted) lines show the predictions for the best-fit $\Lambda$CDM
model from the combined BAO, SNIa, and $H(z)$ data analyses, with
cosmological parameter values $(\Omega_{m0}, \Omega_\Lambda, h) = 
(0.29, 0.72, 0.68) [(0.28, 0.78, 0.738)]$.
}
\label{fig:obs}
\end{figure}

Figures \ref{fig:LCDM_com}---\ref{fig:phiCDM_com} show the
constraints on cosmological parameters for the $\Lambda$CDM and
$\phi$CDM models and the XCDM parametrization, from a joint analysis
of the BAO and SNIa data, as well as from a joint analysis of the
BAO, SNIa, and $H(z)$ data. Including the $H(z)$ data in the analysis 
tightens the constraints, somewhat significantly (sometimes by more 
than two standard deviations), in parts of the parameter spaces. 
Figure \ref{fig:obs} shows the $H(z)$ data and the two best-fit
$\Lambda$CDM models. The $H(z)$ data do support the idea of a 
deceleration to acceleration transition somewhere in the range
$0.5 < z < 1$. 

\begin{table}[t]
\begin{center}
\begin{tabular}{ccc}
\hline\hline
{Model and prior} & {BAO+SNIa} & {BAO+SNIa+$H(z)$} \\
\hline
 {$\Lambda $CDM, $h=0.68 \pm 0.028$} & {0.25 $<$ }$\Omega_{m0}${ $<$ 0.36} & 
      {0.26 $<$ }$\Omega_{m0}${ $<$ 0.33} \\
 {} & {0.53 $<$ }$\Omega _{\Lambda }${ $<$ 0.89} & {0.60 $<$ }$\Omega _{\Lambda }${ $<$ 0.84} \\
 \hline
 {$\Lambda $CDM, $h = 0.738 \pm 0.024$} & {0.25 $<$ }$\Omega_{m0}${ $<$ 0.36} & 
      {0.25 $<$ }$\Omega_{m0}${ $<$ 0.32} \\
 {} & {0.53 $<$ }$\Omega _{\Lambda }${ $<$ 0.89} & {0.66 $<$ }$\Omega _{\Lambda }${ $<$ 0.89} \\
 \hline
 {XCDM, $h = 0.68 \pm 0.028$} & {0.30 $<$ }$\Omega_{m0}${ $<$ 0.38} & 
      {0.27 $<$ }$\Omega_{m0}${ $<$ 0.32} \\
 {} & $-1.18 <  \omega_{\rm X} < -0.78$ & $-1.03 < \omega_{\rm X} < -0.77$ \\
 \hline 
 {XCDM, $h = 0.738 \pm 0.024$} & {0.30 $<$ }$\Omega_{m0}${ $<$ 0.38} & 
      {0.25 $<$ }$\Omega_{m0}${ $<$ 0.30} \\
 {} & $-1.18 < \omega_{\rm X} < -0.78$ & $ -1.15 < \omega_{\rm X} < -0.90$ \\
  \hline
 {$\phi $CDM, $h = 0.68 \pm 0.028$} & {0.25 $<$ }$\Omega_{m0}${ $<$ 0.35} & 
      {0.25 $<$ }$\Omega_{m0}${ $<$ 0.32} \\
 {} & {0 $<$ $\alpha $ $<$ 0.54} & {0 $<$ $\alpha $ $<$ 0.56} \\
  \hline
 {$\phi $CDM, $h = 0.738 \pm 0.024$} & {0.25 $<$ }$\Omega_{m0}${ $<$ 0.35} & 
      {0.25 $<$ }$\Omega_{m0}${ $<$ 0.30} \\
 {} & {0 $<$ $\alpha $ $<$ 0.54} & {0 $<$ $\alpha $ $<$ 0.21} \\
\hline\hline
\end{tabular}

\caption{Two standard deviation bounds on cosmological
parameters using BAO+SNIa and BAO+SNIa+$H(z)$ data, for three 
models and two $H_0$ priors.}
\label{tab:intervals}
\end{center}
\end{table}

Table 1 lists the two standard deviation bounds on the individual 
cosmological parameters, determined from their one-dimensional
posterior probability distributions  functions (which are derived
by marginalizing the two-dimensional likelihood over the other
cosmological parameter). 

\section{Conclusion}
\label{summary}

Adding the \cite{Busca12} $z = 2.3$ measurement of the Hubble parameter,
from BAO in the Ly$\alpha$ forest, to the 21 $H(z)$ data points tabulated
in \cite{Farooq2012}, results in an $H(z)$ data set that provides quite
restrictive constraints on cosmological parameters. These constraints
are tighter than those that follow from the SNIa data of \cite{suzuki12},
which carefully accounts for all known systematic uncertainties. The $H(z)$
field is much less mature than the SNIa one, and there might be some as 
yet undetected $H(z)$ systematic errors that could broaden the $H(z)$ 
error bars, as has happened in the SNIa case. However, we emphasize that 
the observers have done a careful analysis and the error bars we have used 
in our analysis have been carefully estimated. In addition to providing
more restrictive constraints, the $H(z)$ data alone requires accelerated
cosmological expansion at the current epoch at approximately 2 $\sigma$
confidence level, depending on model and $H_0$ prior used in the analysis.  

In summary, the results of the joint analysis of the $H(z)$, BAO, and SNIa 
data are quite consistent with the predictions of the standard spatially-flat
$\Lambda$CDM cosmological model, with current energy budget dominated by 
a time-independent cosmological constant. However, currently-available 
data cannot rule out slowly-evolving dark energy density. We anticipate 
that soon to be available better quality data will more clearly discriminate
between constant and slowly-evolving dark energy density.

\acknowledgments

We thank Data Mania and Larry Weaver for useful discussions and 
helpful advice. This work was supported in part by DOE grant 
DEFG03-99EP41093 and NSF grant AST-1109275.



\begin{thebibliography}{80}
\expandafter\ifx\csname natexlab\endcsname\relax\def\natexlab#1{#1}\fi

\bibitem[{{Allen} {et~al.}(2008)}]{allen08}
{Allen}, S.~W., {et~al.} 2008, MNRAS, 383, 879

\bibitem[{{Anderson} {et~al.}(2012)}]{Anderson2012}
{Anderson}, L., {et~al.} 2012, arXiv:1203.6594 [astro-ph.CO]

\bibitem[{{Astier} \& {Pain}(2012)}]{Astier2012}
Astier, P., \& Pain, R.\ 2012, C.\ R.\ Physique, 13, 521 

\bibitem[{{Aviles} {et~al.}(2012)}]{Aviles2012}
Aviles, A., {et~al.} 2012, arXiv:1204.2007 [astro-ph.CO]

\bibitem[{{Basilakos} {et~al.}(2012)}]{Basilakos2012}
{Basilakos}, S., Polarski, D., \& Sol{\`a}, J.\ 2012, 
arXiv:1204.4806 [astro-ph.CO]

\bibitem[{{Basse} {et~al.}(2012)}]{Basse2012}
{Basse}, T., {et~al.} 2012, arXiv:1205.0548 [astro-ph.CO]

\bibitem[{{Beutler} {et~al.}(2011)}]{beutler2011}
{Beutler}, F., {et~al.} 2011, MNRAS, 416, 3077

\bibitem[{{Blake} {et~al.}(2011)}]{blake11}
{Blake}, C., {et~al.} 2011, MNRAS, 418, 1707

\bibitem[{{Bolotin} {et~al.}(2011)}]{Bolotin2011}
{Bolotin}, Yu.~L., {Lemets}, O.~A., \& Yerokhin, D.~A. 2011, 
arXiv:1108.0203 [astro-ph.CO]

\bibitem[{{Brax} \& {Davis}(2012)}]{Brax2012}
{Brax}, P., \& {Davis}, A.-C. 2012, Phys.\ Lett.\ B, 707, 1

\bibitem[{{Busca} {et~al.}(2012)}]{Busca12}
{Busca}, N. G., {et~al.} 2012, arXiv:1211.2616 [astro-ph]

\bibitem[{{Busti} {et~al.}(2012)}]{Busti2012}
{Busti}, V.~C., {Santos}, R.~C., \& Lima, J.~A.~S. 2012, 
Phys.\ Rev.\ D, 85, 103503

\bibitem[{{Cai} {et~al.}(2012)}]{Cai2012}
{Cai}, R.-G., {at~al.} 2012, Phys.\ Rev.\ D, 86, 023511

\bibitem[{{Calabrese} {et~al.}(2012)}]{calabrese12}
{Calabrese}, E., {et~al.} 2012, Phys.\ Rev.\ D, 86, 043520

\bibitem[{{Campanelli} {et~al.}(2012)}]{campanelli11}
{Campanelli}, L., {et~al.} 2012, Eur.\ Phys.\ J.\ C, 72, 2218 

\bibitem[{{Cao} {et~al.}(2012)}]{Cao2012}
{Cao}, S., {et~al.} 2012, \jcap, 1203, 016

\bibitem[{{Capozziello} \& {De Laurentis}(2011)}]{Capozziello2011}
{Capozziello}, S., \& {De Laurentis}, M. 2011, Phys.\ Rept., 509, 167

\bibitem[{{Chae} {et~al.}(2004)}]{chae04}
{Chae}, K.-H., {et~al.} 2004, \apjl, 607, L71

\bibitem[{{Chen} {et~al.}(2003)}]{Chen2003}
{Chen}, G., {Gott}, J.~R., \& {Ratra}, B. 2003, \pasp, 115, 1269

\bibitem[{{Chen} \& {Ratra}(2003)}]{chen03b}
{Chen}, G., \& {Ratra}, B. 2003, \pasp, 115, 1143

\bibitem[{{Chen} \& {Ratra}(2011a)}]{Chen2011a}
{Chen}, G., \& {Ratra}, B. 2011a, \pasp, 123, 1127

\bibitem[{{Chen} \& {Ratra}(2011b)}]{Chen2011b}
{Chen}, Y., \& {Ratra}, B. 2011b, Phys.\ Lett.\ B, 703, 406

\bibitem[{{Chen} \& {Ratra}(2012)}]{Chen2012}
{Chen}, Y., \& {Ratra}, B. 2012, A{\&}A, 543, A104

\bibitem[{{Colless} {et~al.}(2012)}]{Colless2012}
{Colless}, M., {Beutler}, F., \& {Blake}, C. 2012, 
arXiv:1211.2570 [astro-ph.CO]

\bibitem[{{Costa} {et~al.}(2012)}]{Costa2012}
Costa, F.\ E.\ M., Lima, J.\ A.\ S., \& {Oliveira}, F.\ A.\  2012, 
arXiv:1204.1864 [astro-ph.CO]

\bibitem[{{Dantas} {et~al.}(2011)}]{dantas11}
{Dantas}, M.~A., {et~al.} 2011, Phys. Lett. B, 699, 239

\bibitem[{{Davis} {et~al.}(2007)}]{Davis2007}
{Davis}, T.~M., {et~al.} 2007, \apj, 666, 716

\bibitem[{{De Boni} {et~al.}(2012)}]{DeBoni2012}
{De Boni}, C., {et~al.} 2012, arXiv:1205.3163 [astro-ph.CO]

\bibitem[{{Devi} {et~al.}(2011)}]{Devi2011}
{Devi}, N.~C., {Choudhury}, T.~R., \& {Sen}, A.~A. 2011, 
arXiv:1112.0728 [astro-ph.CO]

\bibitem[{{Duan} {et~al.}(2011)}]{Duan2011}
{Duan}, X., {Li}, Y., \& {Gao}, C. 2011, arXiv:1111.3423 [astro-ph.CO]

\bibitem[{{Farooq} {et~al.}(2012)}]{Farooq2012}
{Farooq}, O., {Mania}, D., \& {Ratra}, B. 2012, arXiv:1211.4253 [astro-ph.CO]

\bibitem[{{Freedman} {et~al.}(2012)}]{Freedman2012}
{Freedman}, W.~L., {et~al.} 2012, \apj, 758, 24

\bibitem[{{Gazta\~{n}aga} {et~al.}(2009)}]{gaztanaga09}
Gazta\~{n}aga, E., Cabr\'{e}, A.,~\& Hui, L. 2009, MNRAS, 399,1663
  
\bibitem[{{Gonzalez} {et~al.}(2012)}]{Gonzalez2012}
{Gonzalez}, A.\ H., {et~al.} 2012, \apj, 753, 163

\bibitem[{{Gott} {et~al.}(2001)}]{Gott2001}
{Gott}, J.~R., {et~al.} 2001, \apj, 549, 1

\bibitem[{{Gu} {et~al.}(2012)}]{Gu2012}
Gu, J.\ A., Lee, C.-C., \& Geng, C.-Q.\ 2012, arXiv:1204.4048 [astro-ph.CO]

\bibitem[{{Hollenstein} {et~al.}(2012)}]{Hollenstein2012}
{Hollenstein}, L., {et~al.} 2012, Phys.\ Rev.\ D, 85, 124031

\bibitem[{{Jackson}(2012)}]{Jackson2012}
{Jackson}, J.\ C.\ 2012, arXiv:1207.0697 [astro-ph.CO]

\bibitem[{{Jassal} {et~al.}(2010)}]{jassal10}
{Jassal}, H.~K., {Bagla}, J.~S., \& {Padmanabhan}, T. 2010, \mnras, 405, 2639

\bibitem[{{Jimenez}(2011)}]{Jimenez2011}
{Jimenez}, R.\ 2011, Fortschr.\ Phys., 59, 602

\bibitem[{{Jimenez} {et~al.}(2003)}]{Jimenezetal2003}
{Jimenez}, R., {et~al.} 2003, \apj, 593, 622

\bibitem[{{Komatsu} {et~al.}(2011)}]{Komatsu2011}
{Komatsu}, E., {et~al.} 2011, \apjs, 192, 18

\bibitem[{{Kumar} (2012)}]{Kumar2012}
{Kumar}, S.\ 2012, \mnras, 422, 2532

\bibitem[{{Lee} \& {Ng}(2007)}]{lee07}
{Lee}, S., \& {Ng}, K.-W. 2007, \prd, 76, 043518

\bibitem[{{Li} {et~al.}(2011a)}]{Li2011a}
 {Li}, M., {et~al.} 2011a, Commun.\ Theor.\ Phys., 56, 525

\bibitem[{Li} {et~al.}(2011b)]{Li2011b} 
{Li}, X.-D., {et~al.} 2011b, \jcap, 1107, 011

\bibitem[{{Lima} \& {Cunha}(2012)}]{Lima2012}
Lima, J.\ A.\ S., \& Cunha, J.\ V.\ 2012, arXiv:1206.0332 [astro-ph.CO]

\bibitem[{{Lu} {et~al.}(2011)}]{Lu2011}
{Lu}, J., {et~al.} 2011, Eur.\ Phys.\ J.\ C, 71, 1800

\bibitem[{{Majerotto} {et~al.}(2012)}]{Majerotto2012}
Majerotto, E., {et~al.} 2012, \mnras, 424, 1392

\bibitem[{{Mania} \& {Ratra}(2012)}]{Mania2012}
{Mania}, D., \& {Ratra}, B. 2012, Phys.\ Lett.\ B, 715, 9

\bibitem[{{Mehta} {et~al.}(2012)}]{Mehta2012}
{Mehta}, K.~T., {et~al.} 2012, arXiv:1202.0092 [astro-ph.CO]

\bibitem[{{Moresco} {et~al.}(2012)}]{moresco12}
{Moresco}, M., {et~al.} 2012, \jcap, 1208, 006

\bibitem[{{Pavlov} {et~al.}(2012)}]{Pavlov2012}
{Pavlov}, A., {Samushia}, L., \& {Ratra}, B. 2012, \apj, 760, 19

\bibitem[{{Peebles}(1984)}]{peebles84}
{Peebles}, P.~J.~E. 1984, \apj, 284, 439

\bibitem[{{Peebles} \& {Ratra}(1988)}]{Peebles&Ratra1988}
{Peebles}, P.~J.~E., \& {Ratra}, B. 1988, \apjl, 325, L17

\bibitem[{{Peebles} \& {Ratra}(2003)}]{Peebles&Ratra2003}
{Peebles}, P.~J.~E., \& {Ratra}, B. 2003, Rev.\ Mod.\ Phys., 75, 559

\bibitem[{{Percival} {et~al.}(2010)}]{Percival2010}
{Percival}, W.~J., {et~al.} 2010, \mnras, 401, 2148

\bibitem[{{Perivolaropoulos}(2010)}]{Perivolaropoulos2010}
{Perivolaropoulos}, L. 2010, J. Phys. Conf. Ser., 222, 012024


\bibitem[{{Plionis} {et~al.}(2010)}]{plionisetal10}
{Plionis}, M., {et~al.} 2010, AIP Conf. Proc. 1241, 267

\bibitem[{{Plionis} {et~al.}(2011)}]{plionisetal11}
{Plionis}, M., {et~al.} 2011, \mnras, 416, 2981

\bibitem[{{Podariu} {et~al.}(2001{\natexlab{a}})}]{podariu01a}
{Podariu}, S., {Nugent}, P., \& {Ratra}, B. 2001{\natexlab{a}}, \apj, 553, 39

\bibitem[{{Podariu} {et~al.}(2001{\natexlab{b}})}]{Podariu2001b}
{Podariu}, S., {et~al.} 2001{\natexlab{b}}, \apj, 559, 9

\bibitem[{{Podariu} \& {Ratra}(2000)}]{podariu2000}
{Podariu}, S., \& {Ratra}, B. 2000, \apj, 532, 109

\bibitem[{Poitras}(2012)]{Poitras2012} 
Poitras, V.\ 2012, \jcap, 1206, 039

\bibitem[{{Ponce de Leon}(2012)}]{PoncedeLeon2012}
Ponce de Leon, J.\ 2012, Class.\ Quant.\ Grav., 29, 135009

\bibitem[{{Ratra}(1991)}]{ratra91}
{Ratra}, B. 1991, \prd, 43, 3802

\bibitem[{{Ratra} \& {Peebles}(1988)}]{Ratra&Peebles1988}
{Ratra}, B., \& {Peebles}, P.~J.~E. 1988, \prd, 37, 3406

\bibitem[{{Ratra} \& {Vogeley}(2008)}]{Ratra08}
{Ratra}, B., \& {Vogeley}, M.~S. 2008, \pasp, 120, 235

\bibitem[{{Riess} {et~al.}(2011)}]{Riess2011}
{Riess}, A.~G., {et~al.} 2011, \apj, 730, 119

\bibitem[{{Ruiz} {et~al.}(2012)}]{Ruiz2012}
{Ruiz}, E.\ J., {et~al.} 2012, \prd, 86, 103004

\bibitem[{{Samushia} {et~al.}(2007)}]{samushia07}
{Samushia}, L., {Chen}, G., \& {Ratra}, B. 2007, arXiv:0706.1963 [astro-ph]

\bibitem[{{Samushia} {et~al.}(2010)}]{Samushiaetal2010}
{Samushia}, L., {et~al.} 2010, Phys. Lett. B, 693, 509

\bibitem[{{Samushia} \& {Ratra}(2006)}]{Samushia&Ratra2006}
{Samushia}, L., \& {Ratra}, B. 2006, \apjl, 650, L5

\bibitem[{{Samushia} \& {Ratra}(2008)}]{Samushia&Ratra2008}
{Samushia}, L., \& {Ratra}, B. 2008, \apjl, 680, L1

\bibitem[{{Samushia} \& {Ratra}(2009)}]{samushia09}
{Samushia}, L., \& {Ratra}, B. 2009, \apj, 701, 1373

\bibitem[{{Samushia} \& {Ratra}(2010)}]{Samushia&Ratra2010}
{Samushia}, L., \& {Ratra}, B. 2010, \apj, 714, 1347

\bibitem[{{Samushia} {et~al.}(2011)}]{Samushia2011}
{Samushia}, L., {et~al.} 2011, \mnras, 410, 1993

\bibitem[{{Sartoris} {et~al.}(2012)}]{Sartoris2012}
{Sartoris}, B., {et~al.} 2012, \mnras, 423, 2503

\bibitem[{{Seikel} {et~al.}(2012)}]{Seikel2012}
Seikel, M., {et~al.} 2012, \prd, 86, 083001

\bibitem[{{Sen} \& {Scherrer}(2008)}]{Sen&Scherrer2008}
{Sen}, A.~A., \& {Scherrer}, R.~J. 2008, Phys. Lett. B, 659, 457

\bibitem[{{Sheykhi} {et~al.}(2012)}]{Sheykhi2012}
{Sheykhi}, A., {et~al.} 2012, Intl.\ J.\ Theo.\  Phys., 51, 1663

\bibitem[{{Simon} {et~al.}(2005)}]{Simon2005}
{Simon}, J., {Verde}, L., \& {Jimenez}, J. 2005, \prd, 71, 123001

\bibitem[{{Solano} \& {Nucamendi}(2012)}]{Solano2012}
Solano, F.\ C., \& Nucamendi, U. 2012, arXiv:1207.0250 [astro-ph.CO]

\bibitem[{{Sorce} {et~al.}(2012)}]{Sorce2012}
{Sorce}, J.~G., Tully, R.~B., \& Courtois, H.~M. 2012, \apj, 758, L12

\bibitem[{{Starkman}(2011)}]{Starkman2011}
{Starkman}, G.~D. 2011, Phil.\ Trans.\ Roy.\ Soc. Lond.\ A, 369, 5018

\bibitem[{{Stern} {et~al.}(2010)}]{Stern2010}
Stern, D., {et~al.} 2010, \jcap 1002, 008

\bibitem[{{Suzuki} {et~al.}(2012)}]{suzuki12}
Suzuki, N., {et~al.} 2012, \apj, 746, 85

\bibitem[{{Tammann} \& {Reindl}(2012)}]{Tammann2012}
{Tammann}, G.~A., \& Reindl, B. 2012, arXiv:1211.4655 [astro-ph.CO]

\bibitem[{{Thakur} {et~al.}(2012)}]{Thakur2012}
{Thakur}, S., {et~al.} 2012, arXiv:1204.2617 [astro-ph.CO]

\bibitem[{{Tong} \& {Noh}(2011)}]{tong11}
{Tong}, M., \& {Noh}, H. 2011, Eur.\ Phys.\ J.\ C, 71, 1586

\bibitem[{{Tonoiu} {et~al.}(2011)}]{Tonoiu2011}
Tonoiu, D., Caramete, A., \& Popa, L.~A.. 2011, Rom.\ Rep.\
Phys.,  63, 879

\bibitem[{{Wang} \& {Dai}(2011)}]{Wang2011}
{Wang}, F.~Y., \& Dai, J.~G. 2011, A{\&}A, 536, A96

\bibitem[{{Wang} \& {Zhang}(2011)}]{WangZhang2012}
{Wang}, H., \& Zhang, T.~J. 2012, \apj, 748, 111

\bibitem[{{Wang}(2012)}]{Wang2012}
{Wang}, Y. 2012, AIP Conf. Proc., 1458, 285

\bibitem[{{Wilson} {et~al.}(2006)}]{wilson06}
{Wilson}, K.~M., {Chen}, G., \& {Ratra}, B. 2006, Mod. Phys. Lett. A, 21, 2197

\bibitem[{{Xu} {et~al.}(2012)}]{Xu2012}
Xu, L., Wang, Y., \& Noh, H.\ 2012, Eur.\ Phys.\ J.\ C, 72, 1931

\bibitem[{{Zhang} \& {Wu}(2010)}]{Zhang2010}
{Zhang}, Q.-J., \& {Wu}, Y.-L. 2010, \jcap, 1008, 038

\end{thebibliography}
\end{document}